\documentclass[sigconf]{acmart}

\setcopyright{acmlicensed}
\copyrightyear{2018}
\acmYear{2018}
\acmDOI{XXXXXXX.XXXXXXX}
\acmConference[Conference acronym 'XX]{Make sure to enter the correct
  conference title from your rights confirmation email}{June 03--05,
  2018}{Woodstock, NY}

\usepackage{multirow}
\usepackage[utf8]{inputenc}
\usepackage{colortbl}
\usepackage{enumitem}
\usepackage[table]{xcolor}

\begin{document}

\title{SIDRec: End-to-End Semantic ID Representation and Modeling for Recommendation}

\author{Zhuohang Jiang}
\authornote{Work done during an internship at Kuaishou Technology.}
\affiliation{%
  \institution{The Hong Kong Polytechnic University}
  \city{Hong Kong SAR}
  \country{China}
}
\email{zhuohang.jiang@connect.polyu.hk}

\author{Yongsen Pan}
\authornotemark[1]
\affiliation{%
  \institution{University of Electronic Science and Technology of China}
  \city{Chengdu, SiChuan}
  \country{China}
}
\email{panys@std.uestc.edu.cn}

\author{Pangjing Wu}
\affiliation{%
  \institution{The Hong Kong Polytechnic University}
  \city{Hong Kong SAR}
  \country{China}
}
\email{pang-jing.wu@connect.polyu.hk}

\author{Zihuai Zhao}
\affiliation{%
  \institution{The Hong Kong Polytechnic University}
  \city{Hong Kong SAR}
  \country{China}
}
\email{zihuai.zhao@connect.polyu.hk}

\author{Zheng Hu}
\authornotemark[1]
\affiliation{%
  \institution{Southwest Jiaotong University}
  \city{Chengdu, SiChuan}
  \country{China}
}
\email{huzheng@swjtu.edu.cn}

\author{Yuxin Chen}
\authornote{Corresponding Author.}
\affiliation{%
  \institution{Kuaishou Technology}
  \city{Beijing}
  \country{China}
}
\email{chenyuxin06@kuaishou.com}

\renewcommand{\shortauthors}{Zhuohang Jiang et al.}

\begin{abstract}
Semantic IDs (SIDs) have emerged as a promising approach for discriminative recommendation by encoding continuous item representations into compact discrete identifiers. However, existing SID-based methods face two fundamental limitations: semantic mismatch between content semantics and collaborative preferences, and objective mismatch between SID construction and downstream recommendation optimization. To address these challenges, we propose SIDRec, an end-to-end recommendation-aware SID learning framework that jointly models content and collaborative information while aligning SID construction with recommendation objectives. Specifically, SIDRec introduces a self-updating fusion mechanism that explicitly integrates pretrained content representations with collaborative embeddings initialized from a trained recommendation model, enabling preference-aware semantic modeling. Furthermore, SIDRec employs an MMoE-inspired joint optimization mechanism with shared experts and task-specific gates, allowing recommendation supervision to directly guide differentiable quantization and produce task-adaptive discrete representations. Extensive experiments on three public benchmark datasets demonstrate that SIDRec consistently outperforms all competing methods across both AUC and GAUC, achieving up to 5.15\% relative improvements over the strongest baselines. Further ablation studies validate the effectiveness and complementarity of collaborative semantic modeling and end-to-end SID optimization.
\end{abstract}

\ccsdesc[500]{Information systems~Recommender systems}

\keywords{Semantic IDs, CTR Prediction, Recommendation Systems.}

\newcommand{\ourname}{SIDRec}

\maketitle
\begin{figure}[t]
    \centering
    \includegraphics[width=\linewidth]{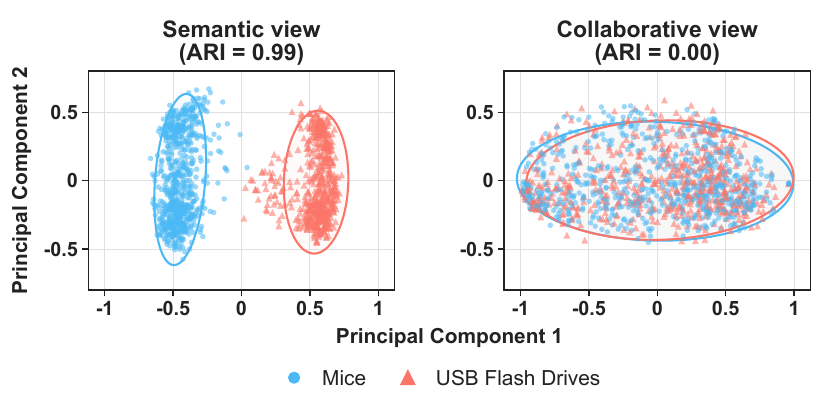}
    \caption{Semantic and collaborative signals capture distinct relationships. Items with different semantics (\emph{e.g.} \textit{mice} and \textit{USB flash drives}) can exhibit similar collaborative patterns.}
    \label{fig:motivation-1}
\end{figure}

\section{Introduction}

Modern recommender systems typically rely on effective item representations to capture user preferences and distinguish among candidate items.
To address this, most existing recommendation models represent each item with an independent identifier learned from historical interactions~\cite{guo2017deepfm, zhou2018din}.
Despite their effectiveness for identifying item-specific patterns, such representations provide limited structure for sharing information across related items, particularly for sparse or newly introduced items.
Recently, Semantic IDs (SIDs) have emerged as a promising alternative by representing items with sequences of discrete semantic codes. 
Specifically, items can be assigned shared codes (e.g., `125' and `129' share codes `1' and `2'), enabling structured knowledge sharing and better generalization than independent item identifiers~\cite{rajput2023recommender,singh2024better}.

Despite these advantages, learning SIDs from content semantics (\emph{e.g.}, what an item is) may not fully capture the collaborative preference patterns (\emph{e.g.}, how users interact with it) in downstream recommendations~\cite{wang2024learnable, ye2025dual}.
For example, two products in the same category with similar descriptions and visual appearances may attract different user groups because of differences in brand preferences, price sensitivity, or intended usage scenarios~\cite{liu2026csg4rec, lin2026token, penha2025semantic}.
Conversely, as illustrated in Figure~\ref{fig:motivation-1}, items with distinct content characteristics may exhibit similar interaction patterns because they are consumed by similar groups of users.
These suggest that item representations for recommendations require both content semantics and collaborative preferences.

\begin{figure*}[t]
\includegraphics[width=\linewidth]{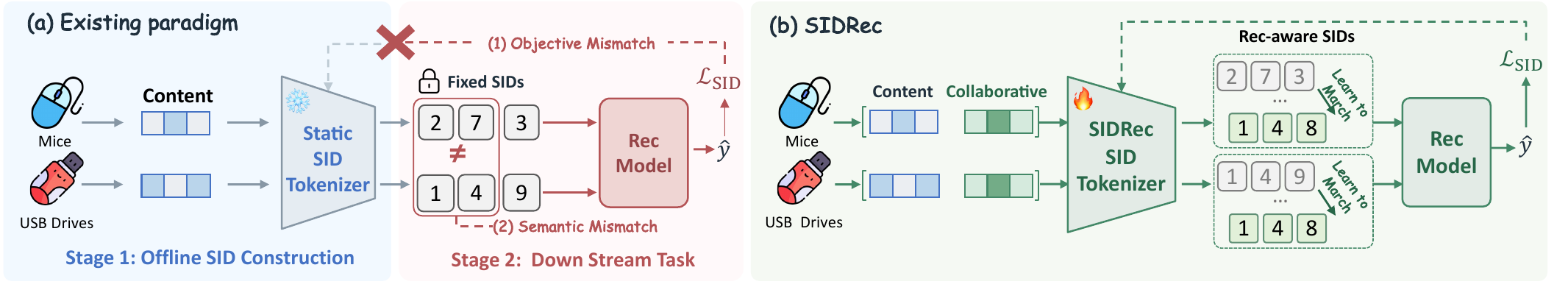}
\caption{Comparison of existing paradigm with our SIDRec. (a) Existing SIDs paradigm mainly rely on content semantics and are optimized independently of downstream recommendation objectives, leading to semantic and objective mismatches. (b) Our SIDRec addresses these issues through content-collaborative semantic modeling and end-to-end SID learning.}
\label{fig:motivation-2}
\end{figure*}

To improve SIDs for recommendations, recent studies have explored incorporating collaborative information into SID learning. 
For example, DIGER~\cite{fu2026differentiable} and DAS~\cite{ye2025dual} adopt exploratory learning and contrastive learning, respectively, to align semantic SIDs and collaborative features.
Meanwhile, DIG~\cite{wang2026discrimination} embeds the SID tokenizer with a ranking model and integrates their losses during training.
However, the collaborative information is primarily introduced through alignment objectives, rather than explicitly fused with content representations. Since quantization is the bottleneck that determines which item representations are preserved in SIDs, when constructing SIDs, the quantization process may not fully carry these two representations.
Moreover, the optimization paradigm for SID learning further compounds such concerns.
In conventional two-stage training, even when collaborative signals are incorporated into SID construction, the resulting SIDs remain fixed during downstream adaptation (i.e., recommendation objectives)~\cite{wang2024learnable, luo2025qarm}. 
As a result, recommendation models can only learn to use SIDs, but cannot update how they are constructed to better support downstream objectives.

These limitations motivate explicitly fusing content and collaborative item representations, and end-to-end SID learning with downstream recommendation objectives, as illustrated in Figure~\ref{fig:motivation-2}.
However, two closely related challenges arise in SID learning.
First, \textbf{semantic mismatch} arises since content semantics and collaborative preferences characterize item representations from fundamentally different perspectives. 
Second, \textbf{objective mismatch} arises since preserving item representations does not necessarily adapt to recommendation objectives.
To address these challenges, we propose SIDRec, an end-to-end SID learning framework that integrates semantic and collaborative information for recommendations. 
To mitigate semantic mismatch, SIDRec introduces a self-updating fusion, which initializes the collaborative embedding table from a trained ranking model and updates it during training.
This allows collaborative representations to evolve with recommendation objectives, capturing interaction patterns that may not be reflected in content semantics alone, while retaining the pretrained semantic embedding of SIDs.

While the self-updating fusion ensures incorporating content and collaborative signals in SIDs, it does not by itself resolve the objective mismatch.
To achieve end-to-end SID adaptation to recommendation objectives, SIDRec draws inspiration from Multi-Gate Mixture of Experts (MMoE) structures~\cite{ma2018modeling} to jointly optimize two objectives. 
Specifically, by sharing experts across the two tasks and using a separate gate for each task, it enables gradients from recommendation objectives to seamlessly propagate through the quantization process of SID learning.
In this manner, the learned SIDs serve not merely as static semantic identifiers, but as recommendation-aware representations tailored to recommendation objectives.

Notably, we conduct extensive experiments on both public benchmarks and real-world industrial datasets. The results demonstrate that SIDRec consistently improves recommendation performance, while comprehensive ablation studies verify the individual contributions of content-collaborative fusion and end-to-end SID learning.
Specifically, SIDRec consistently outperforms all competing methods, achieving up to 5.15\% relative improvement over the strongest baselines. Furthermore, removing either component degrades recommendation performance, with their both removal resulting in relative drops of 7.19\% in AUC and 7.11\% in GAUC, demonstrating the contributions of both components.

The main contributions are summarized as follows:
\begin{itemize}
    \item We identify existing limitations of SIDs for recommendations, including semantic mismatch between content semantics and collaborative preferences, and objective mismatch between SID construction and downstream objectives.

    \item We propose \textbf{SIDRec}, an end-to-end recommendation-aware SID learning framework that addresses these two mismatches by seamlessly integrating content and collaborative information via self-updating, and allowing recommendation objectives to seamlessly optimize SID construction.

    \item We conduct comprehensive experiments on public and industrial recommendation datasets, demonstrating the effectiveness of SIDRec and validating the contributions of SID modeling and corresponding end-to-end learning.
\end{itemize}

\section{Related Work}

\subsection{Semantic IDs for Recommendation}

Conventional recommendation models typically rely on hash-based item identifiers (HIDs), whose embeddings are learned from historical interactions \cite{guo2017deepfm,zhou2018din}. However, these independent identifiers offer limited knowledge sharing across related items, particularly for sparse or newly introduced items \cite{singh2024better,zhu2021learning}. SIDs encode items as sequences of discrete codes, allowing related items to share prefixes rather than unrelated atomic identifiers \cite{rajput2023recommender,singh2024better}. In generative recommendation, TIGER \cite{rajput2023recommender} constructs hierarchical codes through residual quantization for autoregressive item generation. LC-Rec \cite{zheng2024adapting} integrates collaborative semantics into language-model adaptation, and OneRec \cite{deng2025onerec} unifies generative retrieval and ranking. Penha et al. \cite{penha2025semantic} further show that task-specific SIDs do not necessarily transfer between search and recommendation, highlighting the influence of the underlying embedding space. For discriminative ranking, Singh et al. \cite{singh2024better} use content-derived SIDs to generalize to new and long-tail items, while Zheng et al. \cite{zheng2025enhancing} improve embedding stability with prefix-based parameterization. SID-Coord \cite{li2026sidcoord} balances SID generalization and hashed-ID memorization, and QARM \cite{luo2025qarm} learns downstream embeddings for quantized multimodal representations in retrieval and ranking. However, prefix sharing also introduces collisions, which AdaSID \cite{pan2026beyond} addresses through adaptive overlap regulation. In these ranking applications, learning token embeddings does not necessarily update the code assignments, which are commonly determined before downstream optimization.

\subsection{Learnable Semantic IDs}

Despite their advantages, conventional SIDs are often constructed offline from content representations and remain fixed during downstream recommendation training \cite{rajput2023recommender,singh2024better}. This paradigm may overlook collaborative preferences and prevents recommendation objectives from directly refining SID assignments. Recent studies incorporate interaction evidence into SID construction in different ways. TokenRec \cite{qu2025tokenrec} quantizes collaborative embeddings, whereas LETTER \cite{wang2024learnable} regularizes content-based codes through collaborative contrastive learning and code diversity. DAS \cite{ye2025dual} jointly optimizes quantization and multi-view collaborative alignment, while CSG4Rec \cite{liu2026csg4rec} uses collaborative similarity to organize quantized item and user representations for generative sequential recommendation. These approaches distinguish direct use of interaction-derived inputs from supervision that shapes code-space geometry. However, preserving collaborative neighborhoods does not guarantee that quantization boundaries favor candidate-level click discrimination. Another research direction concerns how tokenizers interact with recommendation losses. ETEGRec \cite{liu2025generative} alternates tokenizer and generative-model training, while DIGER \cite{fu2026differentiable} employs differentiable assignments and uncertainty-controlled exploration under generative objectives. DIG \cite{wang2026discrimination} embeds a tokenizer inside a discriminative ranker, allowing ranking feedback to influence codebook boundaries while also supporting retrieval. In contrast, SIDRec explicitly integrates content and collaborative representations through self-updating fusion to mitigate semantic mismatch. It further enables recommendation objectives to guide differentiable SID learning through joint optimization, addressing the objective mismatch between SID construction and downstream recommendation.

\section{Methodology}
\label{sec:method}

\begin{figure*}[t]
    \centering
    \includegraphics[width=\linewidth]{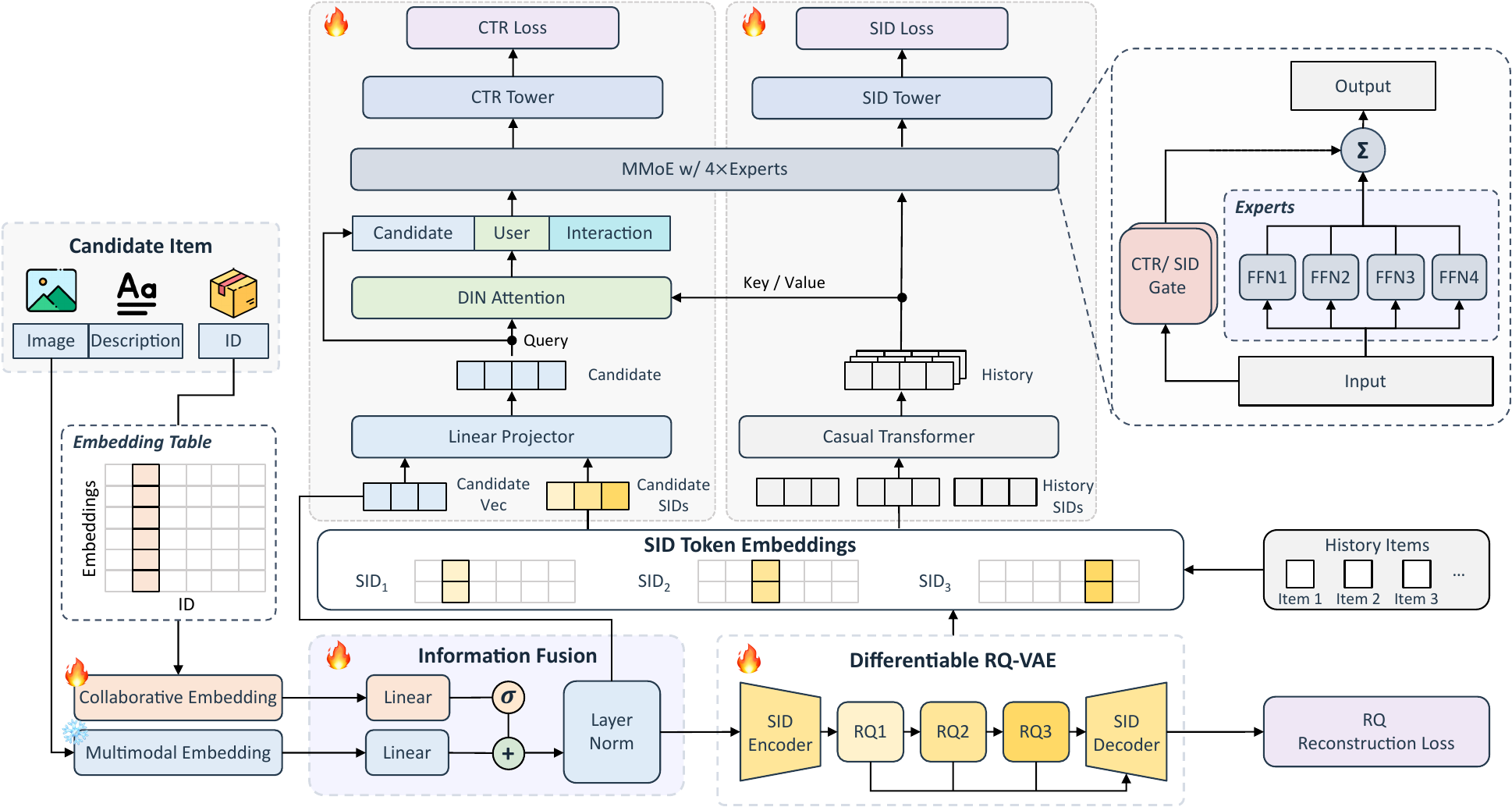}
    \caption{Overview of the SIDRec framework. It first integrates collaborative representations from a trained ranking model with pretrained multimodal embeddings for SID construction. The framework then adapts a pretrained SID tokenizer through downstream CTR learning. Candidate and history representations built from the learned SIDs are used for click prediction and SID prediction, which are optimized jointly.}
    \label{fig:framework}
\end{figure*}

In this section, we present SIDRec to enable end-to-end SID learning, which integrates both semantic and collaborative information for recommendations, as illustrated in Figure~\ref{fig:framework}. 

\par\medskip\noindent\textbf{Notation.} For a user $u$, let $\mathcal H=(i_1,\ldots,i_T)$ denote the interaction history before candidate item $i$. The model predicts the click probability $\hat y\in(0,1)$ for the label $y\in\{0,1\}$. Each item has a fixed multimodal embedding $\mathbf m_i\in\mathbb R^{d_m}$ and a collaborative embedding $\mathbf e_i\in\mathbb R^{d_c}$. We use $d_z$ and $d$ for the tokenizer and ranking dimensions. Each item is represented by an $L$-level SID $\mathbf s_i=(s_i^{(1)},\ldots,s_i^{(L)})$, where $s_i^{(\ell)}\in\{1,\ldots,K\}$ is the token at level $\ell$. We use $L=3$ in SIDRec.

\subsection{Self-Updating Collaborative--Multimodal Fusion}
\label{sec:method:fusion}
Existing SID construction often relies on item content and does not directly model the interaction patterns captured by collaborative representations. SIDRec combines these two sources before quantization. We initialize the collaborative embedding table $\mathbf E\in\mathbb R^{|\mathcal I|\times d_c}$ from a trained ranking model, where $\mathcal I$ is the item set and $\mathbf e_i$ is its $i$-th row. The table continues to be updated during joint training, while the multimodal embeddings remain fixed.

We project both representations into a shared space and control the collaborative contribution with a learnable sigmoid gate. The fused representation is
\begin{equation}
\label{eq:fusion}
\mathbf f_i=\operatorname{LayerNorm}\!\left(\mathbf W_m\mathbf m_i+\sigma(a)\mathbf W_e\mathbf e_i\right),
\end{equation}
where $\sigma(\cdot)$ is the sigmoid function, $a$ is a trainable scalar, and $\mathbf W_m\in\mathbb R^{d_z\times d_m}$ and $\mathbf W_e\in\mathbb R^{d_z\times d_c}$ are projection matrices. The result $\mathbf f_i\in\mathbb R^{d_z}$ is passed to the SID tokenizer. This step makes both item content and interaction-derived preferences available for SID learning. It is expected to help the resulting codes retain preference-relevant item differences for CTR prediction.

\subsection{End-to-End SID Adaptation}
\label{sec:method:sid}
In conventional SID-enhanced ranking, item codes are often learned offline and then fixed, so ranking feedback cannot refine their assignments. SIDRec starts from a pretrained tokenizer and updates its encoder and codebooks together with the ranking model.

The SID encoder first produces $\mathbf z_i=\operatorname{Enc}(\mathbf f_i)\in\mathbb R^{d_z}$. The residual quantizer then selects a codeword at each level to represent the remaining information. Starting from $\mathbf r_i^{(1)}=\mathbf z_i$, the selected codeword and the next residual are
\begin{equation}
\label{eq:rq}
\begin{aligned}
\mathbf c_i^{(\ell)}&=\underset{\mathbf c\in\mathbf C^{(\ell)}}{\arg\min}\,\|\mathbf r_i^{(\ell)}-\mathbf c\|_2^2,\\
\mathbf r_i^{(\ell+1)}&=\mathbf r_i^{(\ell)}-\mathbf c_i^{(\ell)},
\end{aligned}
\end{equation}
where $\mathbf C^{(\ell)}\in\mathbb R^{K\times d_z}$ is the codebook at level $\ell$ and $\mathbf c_i^{(\ell)}\in\mathbb R^{d_z}$ is its selected row. The corresponding row index is $s_i^{(\ell)}$. The quantized representation $\hat{\mathbf z}_i=\sum_{\ell=1}^{L}\mathbf c_i^{(\ell)}$ is decoded as $\hat{\mathbf m}_i=\operatorname{Dec}(\hat{\mathbf z}_i)$.

To pass CTR gradients to the tokenizer, we use straight-through quantization. Each item uses a hard SID code in the forward pass, while gradients follow Gumbel-perturbed soft codeword probabilities. Let $\tilde{\mathbf p}_i^{(\ell)}\in\mathbb R^K$ denote this assignment. Its SID embedding is
\begin{equation}
\label{eq:token}
\mathbf b_i^{(\ell)}=\bigl(\mathbf B^{(\ell)}\bigr)^\top\tilde{\mathbf p}_i^{(\ell)},
\end{equation}
where $\mathbf B^{(\ell)}\in\mathbb R^{K\times d}$ is the trainable SID embedding table at level $\ell$. This gives the embedding of the selected code during the forward pass and carries CTR gradients back to the tokenizer.

The candidate branch projects $\mathbf z_i$ together with the SID embeddings $\mathbf b_i^{(1)},\ldots,\mathbf b_i^{(L)}$ into a candidate representation $\mathbf q_i\in\mathbb R^d$. The history branch uses the same token embeddings to model interactions over time. Let $\mathbf X\in\mathbb R^{TL\times d}$ contain the SID embeddings of $\mathcal H$ in temporal order and let $\mathbf P$ denote positional embeddings. The causal Transformer applies masked self-attention and feed-forward transformations to obtain
\begin{equation}
\label{eq:causal}
\mathbf G=\operatorname{FFN}\!\left(\operatorname{Attn}(\mathbf X+\mathbf P)\right),
\end{equation}
where $\mathbf G\in\mathbb R^{TL\times d}$ contains the contextual token states. Let $\mathbf h_t^{(\ell)}$ denote the state for level $\ell$ of history item $i_t$. We obtain its item-level representation by $\mathbf h_t=\frac{1}{L}\sum_{\ell=1}^{L}\mathbf h_t^{(\ell)}$ and a candidate-independent history summary by $\bar{\mathbf h}=\frac{1}{T}\sum_{t=1}^{T}\mathbf h_t$, with pooling over valid history positions.

For CTR prediction, we use DIN attention to summarize history in relation to the candidate item. With $\mathbf H=[\mathbf h_1,\ldots,\mathbf h_T]$, the candidate-aware user representation is
\begin{equation}
\label{eq:din}
\mathbf u_i=\operatorname{Attn}(\mathbf q_i,\mathbf H),
\end{equation}
where $\operatorname{Attn}$ denotes DIN target attention, which scores each history item using the candidate, history representation, their difference, and their elementwise product. The resulting $\mathbf u_i\in\mathbb R^d$ captures the historical interests relevant to the current candidate. Together, the history and candidate branches let the CTR task update SID construction using preference signals from both sides of the interaction.

\subsection{Joint Optimization}
\label{sec:method:joint}
SIDRec learns click prediction together with an auxiliary SID prediction task. The CTR task uses candidate-aware interests, while SID prediction uses the history summary to predict the codes of the positive target item. We employ a Multi-Gate Mixture of Experts (MMoE) to jointly optimize these two objectives by sharing four experts across the two tasks and using a separate gate for each task.

The CTR input $\mathbf x_c\in\mathbb R^{3d}$ combines $\mathbf u_i$, $\mathbf q_i$, and their elementwise product $\mathbf u_i \odot \mathbf q_i$. The SID input $\mathbf x_s\in\mathbb R^{3d}$ contains $\bar{\mathbf h}$ with the remaining dimensions padded. For $r\in\{c,s\}$, task-specific gating gives
\begin{equation}
\label{eq:mmoe}
\begin{aligned}
\mathbf g_r&=\operatorname{softmax}(\mathbf W_{g,r}\mathbf x_r),\\
\mathbf o_r&=\mathbf g_r\operatorname{FFN}(\mathbf x_r),
\end{aligned}
\end{equation}
where $\mathbf W_{g,r}$ is the task-specific gating matrix, $\mathbf g_r\in\mathbb R^4$ contains the expert weights, and $\mathbf o_r$ is the representation passed to task $r$.

The CTR tower maps its task representation to the predicted click probability
\begin{equation}
\label{eq:ctrhead}
\hat y=\sigma\!\left(\operatorname{FFN}_c(\mathbf o_c)\right),
\end{equation}
where $\operatorname{FFN}_c$ is the CTR prediction tower. The SID tower predicts a distribution over the $K$ codes at each quantization level
\begin{equation}
\label{eq:sidhead}
\hat{\mathbf p}^{(\ell)}=\operatorname{softmax}\!\left(\mathbf W_s^{(\ell)}\mathbf o_s\right),
\end{equation}
where $\hat{\mathbf p}^{(\ell)}\in\mathbb R^K$ is conditioned on the observed history.

The CTR loss is computed over all labeled training examples
\begin{equation}
\label{eq:ctr}
\mathcal L_{\mathrm{ctr}}=\frac{1}{N}\sum_{n=1}^{N}\operatorname{BCE}(y_n,\hat y_n),
\end{equation}
where $N$ denotes the total number of training examples and $\operatorname{BCE}$ is binary cross-entropy. 

The SID prediction loss measures how well the history representation predicts the positive target's SID at every level
\begin{equation}
\label{eq:sidpred}
\mathcal L_p=-\frac{1}{ML}\sum_{n=1}^{M}\sum_{\ell=1}^{L}\log\hat p_n^{(\ell)},
\end{equation}
where $M$ denotes the number of positive examples and $\hat p_n^{(\ell)}$ is the probability assigned to the observed target code at level $\ell$. These codes are supplied by the current tokenizer and used as fixed labels within each update. The quantization objective also reconstructs the multimodal input to preserve item information during SID adaptation. For $R$ reconstructed items, it is
\begin{equation}
\label{eq:rq_loss}
\mathcal L_q=\frac{1}{Rd_m}\sum_{j=1}^{R}\|\mathbf m_j-\hat{\mathbf m}_j\|_2^2+\mathcal L_{\mathrm{vq}},
\end{equation}
where $\mathcal L_{\mathrm{vq}}$ includes the codebook and commitment penalties of residual quantization. The SID learning objective combines target-code prediction with quantization regularization
\begin{equation}
\label{eq:sid_loss}
\mathcal L_{\mathrm{sid}}=\lambda_p\mathcal L_p+\lambda_q\mathcal L_q,
\end{equation}
where $\lambda_p$ and $\lambda_q$ control their relative contributions.

Finally, CTR prediction and SID learning are optimized through a unified objective
\begin{equation}
\label{eq:total}
\mathcal L=\mathcal L_{\mathrm{ctr}}+\mathcal L_{\mathrm{sid}}.
\end{equation}
This objective lets click prediction guide SID adaptation through the differentiable tokenizer while the SID learning terms support target-code prediction and multimodal reconstruction. The shared experts also allow both tasks to learn from the same history representations.

\section{Experiments}
\label{sec:experiments}

In this section, we evaluate SIDRec on public datasets and in real-world industrial scenarios to assess its effectiveness for fine ranking. Our experiments address three research questions.
\begin{itemize}
    \setlength{\itemsep}{1pt}
    \setlength{\parskip}{0pt}
    \item \textbf{RQ1.} How does SIDRec perform on public benchmarks and across user activity levels?
    \item \textbf{RQ2.} How do collaborative information and end-to-end optimization affect ranking performance and the learned SID structure?
    \item \textbf{RQ3.} How does SIDRec perform across industrial domains and prediction tasks?
\end{itemize}

\subsection{Experimental Settings}
\label{sec:exp_settings}

\subsubsection{Datasets}

\noindent\textit{Public datasets.}
We use \textbf{Amazon Electronics}, \textbf{Amazon Beauty}, and \textbf{MovieLens-20M}, covering product and movie recommendation. The Amazon datasets use the 5-core review data. For MovieLens-20M, ratings of at least 4 are treated as positive interactions. Table~\ref{tab:dataset_statistics} reports the processed dataset statistics. Each item has a fixed 4,096-dimensional semantic representation.

The public preprocessing pipeline sorts each user's interactions by time. For eligible users, the last interaction is held out for testing and the preceding one for validation. Earlier interactions form history--target training pairs. We evaluate candidate-level binary prediction with sampled negatives. All compared methods use the same data splits and evaluation candidate sets.
\begin{table}[t]
    \centering
    \caption{Statistics of the processed public datasets. History lengths are measured before truncation.}
    \label{tab:dataset_statistics}
    \small
    \setlength{\tabcolsep}{3pt}
    \begin{tabular*}{\linewidth}{@{\extracolsep{\fill}}lrrr@{}}
        \toprule
        \textbf{Statistic} & \textbf{Electronics} & \textbf{ML-20M} & \textbf{Beauty} \\
        \midrule
        Users & 192,403 & 138,287 & 22,363 \\
        Items & 63,001 & 20,720 & 12,101 \\
        Interactions & 1,689,188 & 9,995,410 & 198,502 \\
        Average history & 8.78 & 72.28 & 8.88 \\
        Maximum history & 431 & 3,177 & 204 \\
        Sparsity & 99.986\% & 99.651\% & 99.927\% \\
        \bottomrule
    \end{tabular*}
\end{table}

\noindent\textit{Industrial datasets.}
We also evaluate SIDRec on two domains of a large-scale recommendation platform, denoted as \textbf{Domain 1} and \textbf{Domain 2}. Each domain includes three prediction targets, namely exposure (\textit{Show}), clicks (\textit{CTR}), and conversions (\textit{CVR}). These domains provide real-world interaction data for evaluating SIDRec across multiple industrial prediction objectives.

\subsubsection{Baselines}
We compare SIDRec against fifteen representative CTR models, grouped according to their primary modeling mechanisms.
\begin{itemize}[
    leftmargin=2em,
    labelsep=0.4em,
    itemsep=2pt,
    parsep=0pt,
    topsep=2pt
]
    \item \textit{Linear and factorized models.} \textbf{LR} models additive feature effects.
    \textbf{FM}~\cite{rendle2010fm} captures pairwise interactions through factorized parameters.

    \item \textit{Feature interaction networks.} \textbf{DeepFM}~\cite{guo2017deepfm} combines factorization machines with a deep network.
    \textbf{DCNv2}~\cite{wang2021dcnv2} combines explicit feature crosses with nonlinear transformations.
    \textbf{FinalMLP}~\cite{mao2023finalmlp} uses two MLP streams with feature selection and bilinear fusion.
    \textbf{GDCN}~\cite{wang2023gdcn} introduces gated cross layers to filter high-order interactions.
    \textbf{WuKong}~\cite{zhang2024wukong} stacks factorization-machine layers to capture high-order interactions.
    \textbf{RankMixer}~\cite{zhu2025rankmixer} adopts efficient token mixing to scale feature interactions.
    \textbf{QNN}~\cite{li2025qnn} models feature interactions using quadratic neural transformations.

    \item \textit{Personalized networks.} \textbf{PPNet}~\cite{chang2023pepnet} uses personalized gates to modulate deep-network parameters.

    \item \textit{Interest and behavior models.} \textbf{DIN}~\cite{zhou2018din} uses target-aware attention to aggregate historical behaviors.
    \textbf{DIEN}~\cite{zhou2019dien} models the temporal evolution of user interests.
    \textbf{TransAct}~\cite{xia2023transact} uses a Transformer to capture short-term preferences from real-time user actions.

    \item \textit{Unified sequence--interaction models.} \textbf{OneTrans}~\cite{zhang2026onetrans} jointly models feature interactions and behavior sequences in a single Transformer.
    \textbf{HyFormer}~\cite{huang2026hyformer} integrates long-sequence modeling and heterogeneous feature interactions in a hybrid Transformer.
\end{itemize}

\subsubsection{Evaluation Metrics}
For public datasets, we report \textbf{AUC} and \textbf{GAUC}. AUC measures discrimination over all evaluated examples. GAUC averages per-user AUC scores with weights proportional to each user's number of examples. Users with only one label class are excluded from GAUC. Higher AUC and GAUC indicate better performance. Industrial evaluation reports AUC and the platform's user-level metrics, \textbf{UAUC} and \textbf{WUAUC}.

Relative gains in the public comparison are computed against the corresponding baseline for each metric. Component ablation reports relative percentage drops (\%), while industrial gains are expressed in percentage points (pp).

\definecolor{gain0}{RGB}{248,249,248}
\definecolor{gain1}{RGB}{239,245,241}
\definecolor{gain2}{RGB}{224,238,228}
\definecolor{gain3}{RGB}{205,227,212}
\definecolor{gain4}{RGB}{181,215,191}
\definecolor{gain5}{RGB}{151,199,164}
\definecolor{lightgray}{RGB}{242,242,242}

\newcommand{\gaincell}[2]{%
  \cellcolor{gain#1}\textcolor{black}{+#2\%}%
}

\begin{table*}[t]
    \centering
    \caption{Performance comparison on three public datasets. Each $\Delta$ column reports the relative improvement (\%) of SIDRec over the corresponding baseline. Darker green cells indicate larger gains within each metric, and all gain values are printed in black. Bold denotes the best result and underlining denotes the strongest baseline. Rank is determined by the mean of the six per-metric ranks. Equal mean ranks share the same rank.}
    \label{tab:ctr_main}
    \setlength{\tabcolsep}{2pt}
    \renewcommand{\arraystretch}{1.09}
    \resizebox{\textwidth}{!}{%
    \begin{tabular}{ll*{12}{r}r}
        \toprule
        \multirow{2}{*}{\textbf{Category}} & \multirow{2}{*}{\textbf{Model}}
        & \multicolumn{4}{c}{\textbf{Electronics}}
        & \multicolumn{4}{c}{\textbf{MovieLens-20M}}
        & \multicolumn{4}{c}{\textbf{Beauty}}
        & \multirow{2}{*}{\textbf{Rank}} \\
        \cmidrule(lr){3-6}\cmidrule(lr){7-10}\cmidrule(lr){11-14}
        & & \textbf{AUC} & $\boldsymbol{\Delta}$ & \textbf{GAUC} & $\boldsymbol{\Delta}$
          & \textbf{AUC} & $\boldsymbol{\Delta}$ & \textbf{GAUC} & $\boldsymbol{\Delta}$
          & \textbf{AUC} & $\boldsymbol{\Delta}$ & \textbf{GAUC} & $\boldsymbol{\Delta}$ & \\
        \midrule
        \multirow{2}{*}{Linear} & LR & 0.75960 & \gaincell{5}{15.79} & 0.75912 & \gaincell{5}{15.60} & 0.94158 & \gaincell{4}{4.40} & 0.94143 & \gaincell{5}{4.09} & 0.66963 & \gaincell{5}{23.02} & 0.66640 & \gaincell{5}{20.97} & 16 \\
         & FM & 0.76155 & \gaincell{5}{15.49} & 0.76176 & \gaincell{5}{15.20} & 0.92802 & \gaincell{5}{5.93} & 0.93990 & \gaincell{5}{4.26} & 0.72163 & \gaincell{3}{14.15} & 0.70956 & \gaincell{3}{13.61} & 11 \\
        \midrule
        \multirow{7}{*}{Interaction} & DeepFM & 0.76340 & \gaincell{5}{15.21} & 0.75985 & \gaincell{5}{15.49} & 0.94452 & \gaincell{3}{4.08} & 0.94517 & \gaincell{4}{3.67} & 0.71900 & \gaincell{3}{14.57} & 0.70124 & \gaincell{3}{14.96} & 13 \\
         & DCNv2 & 0.76428 & \gaincell{5}{15.08} & 0.76084 & \gaincell{5}{15.34} & 0.94725 & \gaincell{3}{3.78} & 0.94884 & \gaincell{4}{3.27} & 0.72021 & \gaincell{3}{14.38} & 0.70647 & \gaincell{3}{14.11} & 8 \\
         & FinalMLP & 0.76283 & \gaincell{5}{15.30} & 0.76035 & \gaincell{5}{15.42} & 0.94484 & \gaincell{3}{4.04} & 0.94480 & \gaincell{4}{3.72} & 0.71699 & \gaincell{3}{14.89} & 0.70083 & \gaincell{3}{15.03} & 14 \\
         & GDCN & 0.76418 & \gaincell{5}{15.10} & 0.76132 & \gaincell{5}{15.27} & 0.94685 & \gaincell{3}{3.82} & 0.94694 & \gaincell{4}{3.48} & 0.72321 & \gaincell{3}{13.90} & 0.70681 & \gaincell{3}{14.05} & 8 \\
         & WuKong & 0.76471 & \gaincell{5}{15.02} & 0.76262 & \gaincell{5}{15.07} & 0.94832 & \gaincell{3}{3.66} & 0.94903 & \gaincell{4}{3.25} & 0.70024 & \gaincell{4}{17.64} & 0.68648 & \gaincell{4}{17.43} & 10 \\
         & RankMixer & 0.82956 & \gaincell{1}{6.02} & 0.82776 & \gaincell{1}{6.02} & 0.97298 & \gaincell{1}{1.03} & 0.96928 & \gaincell{1}{1.10} & 0.76518 & \gaincell{1}{7.66} & 0.74149 & \gaincell{1}{8.72} & 5 \\
         & QNN & 0.75839 & \gaincell{5}{15.97} & 0.75718 & \gaincell{5}{15.90} & 0.94641 & \gaincell{3}{3.87} & 0.94461 & \gaincell{4}{3.74} & 0.68115 & \gaincell{4}{20.94} & 0.67123 & \gaincell{5}{20.10} & 15 \\
        \midrule
        \multirow{1}{*}{Personalized} & PPNet & 0.76228 & \gaincell{5}{15.38} & 0.75994 & \gaincell{5}{15.48} & 0.94589 & \gaincell{3}{3.93} & 0.94810 & \gaincell{4}{3.35} & 0.71630 & \gaincell{3}{15.00} & 0.70215 & \gaincell{3}{14.81} & 11 \\
        \midrule
        \multirow{3}{*}{Interest} & DIN & 0.81627 & \gaincell{2}{7.75} & 0.81516 & \gaincell{2}{7.66} & 0.97206 & \gaincell{1}{1.13} & 0.96797 & \gaincell{1}{1.23} & 0.74149 & \gaincell{2}{11.10} & 0.72598 & \gaincell{2}{11.04} & 7 \\
         & DIEN & 0.83576 & \gaincell{1}{5.24} & 0.83408 & \gaincell{1}{5.22} & 0.97318 & \gaincell{1}{1.01} & 0.96791 & \gaincell{1}{1.24} & 0.75811 & \gaincell{1}{8.66} & 0.73620 & \gaincell{1}{9.50} & 5 \\
         & TransAct & \underline{0.85154} & \gaincell{0}{3.29} & \underline{0.85045} & \gaincell{0}{3.19} & \underline{0.97989} & \gaincell{0}{0.32} & \underline{0.97573} & \gaincell{0}{0.43} & \underline{0.79001} & \gaincell{0}{4.27} & \underline{0.76663} & \gaincell{0}{5.15} & 2 \\
        \midrule
        \multirow{2}{*}{Unified} & OneTrans & 0.84125 & \gaincell{0}{4.55} & 0.84002 & \gaincell{1}{4.47} & 0.97605 & \gaincell{0}{0.71} & 0.97269 & \gaincell{0}{0.74} & 0.77914 & \gaincell{0}{5.73} & 0.75543 & \gaincell{0}{6.71} & 3 \\
         & HyFormer & 0.83535 & \gaincell{1}{5.29} & 0.83556 & \gaincell{1}{5.03} & 0.97534 & \gaincell{0}{0.79} & 0.97125 & \gaincell{1}{0.89} & 0.77842 & \gaincell{0}{5.82} & 0.75616 & \gaincell{0}{6.61} & 4 \\
        \midrule
        \rowcolor{lightgray}
        \multirow{1}{*}{\textbf{Ours}} & \textbf{SIDRec} & \textbf{0.87954} &  & \textbf{0.87758} &  & \textbf{0.98302} &  & \textbf{0.97990} &  & \textbf{0.82376} &  & \textbf{0.80614} &  & \textbf{1} \\
        \bottomrule
    \end{tabular}%
    }
\end{table*}

\subsubsection{Implementation Details}
We evaluate SIDRec across diverse public benchmarks, industrial scenarios, and ranking objectives.

\noindent\textit{General backbone.}
On both public benchmarks and industrial scenarios, SIDRec uses a causal Transformer to encode historical SID tokens and a DIN attention module to extract candidate-aware interests. The candidate representation combines the tokenizer latent with embeddings of its SID codes. Four shared MMoE experts with task-specific gates support CTR prediction and auxiliary prediction of the positive target's SID. The SID tokenizer combines semantic--collaborative fusion, an MLP encoder, residual quantization, and a reconstruction decoder. During CTR training, the pretrained semantic tables remain fixed, while the collaborative embedding, fusion projections, tokenizer, and ranking network are finetuned.

\noindent\textit{General reference configuration.}
The code configures three residual codebooks, each containing 256 codewords, with a latent dimension of 256. The history encoder has two Transformer layers, a hidden size of 128, four attention heads, a feed-forward width of 512, and dropout of 0.1. Histories are truncated to the most recent 50 items. The DIN attention width is 64, and the expert and prediction-tower hidden widths are 128. The optimizer is AdamW, with reference learning rates of $10^{-3}$ for the ranker and $10^{-5}$ for the tokenizer during joint training. Checkpoint selection uses validation GAUC. The reference CTR protocol samples 10 negatives per positive for training and 100 for evaluation. Its MovieLens configuration caps training at 1,000,000 history--target pairs and leaves evaluation uncapped. The reported public results use seed 2026.

\noindent\textit{Industrial setup.}
In real-world industrial scenarios, we evaluate SIDRec against the corresponding industrial baseline for each domain and each of the \textit{Show}, \textit{CTR}, and \textit{CVR} targets. Performance is assessed through offline changes in prediction metrics. 

\subsection{Performance on Public Datasets (RQ1)}
\label{sec:exp_public}
We first evaluate SIDRec against representative CTR baselines on three public datasets and examine its performance across different user activity levels.

\subsubsection{Overall Comparison}
Table~\ref{tab:ctr_main} compares SIDRec with fifteen baselines. The results reveal two main patterns.
\begin{itemize}
    \setlength{\itemsep}{2pt}
    \setlength{\parskip}{0pt}
    \item \textbf{Overall ranking quality.} SIDRec achieves the highest AUC and GAUC on all three datasets and ranks first by the mean of the six per-metric ranks. TransAct is the strongest baseline for all six metrics and ranks second overall, followed by OneTrans and HyFormer in third and fourth place. Relative AUC gains over TransAct are 3.29\%, 0.32\%, and 4.27\% on Electronics, MovieLens-20M, and Beauty, respectively. The corresponding GAUC gains are 3.19\%, 0.43\%, and 5.15\%. These improvements indicate stronger global discrimination and per-user ranking quality.
    \item \textbf{Dataset differences.} The largest relative gains in both metrics occur on Beauty, followed by Electronics and MovieLens-20M. Both Amazon datasets have shorter average interaction histories than MovieLens-20M. On MovieLens-20M, SIDRec improves AUC from 0.97989 to 0.98302 and GAUC from 0.97573 to 0.97990.
\end{itemize}
These results show that SIDRec outperforms all fifteen CTR baselines in both AUC and GAUC across the three public benchmarks.

\begin{table}[t]
    \centering
    \caption{Performance of SIDRec-CTR across different user activity levels on the Beauty dataset.}
    \label{tab:user_activity}
    \begin{tabular}{p{80pt}ccc}
        \toprule
        \textbf{History Length}
        & \textbf{Users (\%)}
        & \textbf{AUC}
        & \textbf{GAUC} \\
        \midrule
        3--5   & 11,383 (50.9\%) & 0.8121 & 0.7973 \\
        6--10  & 7,433 (33.2\%)  & 0.8213 & 0.8044 \\
        11--20 & 2,528 (11.3\%)  & 0.8533 & 0.8281 \\
        21--50 & 875 (3.9\%)     & 0.8849 & 0.8597 \\
        51+    & 144 (0.6\%)     & \textbf{0.9189} & \textbf{0.8846} \\
        \midrule
        \rowcolor{lightgray}\textit{Overall} & 22,363 (100\%) & 0.8238 & 0.8061 \\
        \bottomrule
    \end{tabular}
\end{table}

\subsubsection{User Activity}
We group Beauty users by their historical interaction lengths to examine how performance varies with user activity. Table~\ref{tab:user_activity} shows two main patterns.

\begin{itemize}
    \setlength{\itemsep}{2pt}
    \setlength{\parskip}{0pt}
    \item \textbf{History-length trend.} AUC and GAUC increase across the five groups, from 0.8121 and 0.7973 for users with 3--5 interactions to 0.9189 and 0.8846 for users with more than 50 interactions. This trend indicates that SIDRec effectively leverages richer interaction histories to model user interests.
    \item \textbf{User coverage.} Users with 3--10 interactions account for 84.1\% of the evaluated users, whereas the longest-history group contains 144 users. The concentration of users in the short-history groups highlights the importance of effective ranking for users with limited interactions.
\end{itemize}
These results show that SIDRec achieves effective ranking across different activity levels, including the short-history users who constitute the majority of the Beauty dataset.

\subsection{Analysis of SID Learning (RQ2)}
\label{sec:exp_sid_analysis}
We then conduct ablation studies to assess the contributions of collaborative fusion and end-to-end optimization, followed by analyses of how collaborative signals affect SID assignments and their structural consistency.

\subsubsection{Component Ablation}
\label{sec:exp_ablation}

We conduct an ablation study on Electronics to investigate the contributions of collaborative fusion and end-to-end SID optimization. Specifically, \textit{w/o Collaborative View} removes collaborative information from SID construction, \textit{w/o End-to-end SID} freezes SID construction during ranking training, and \textit{w/o Both} removes both components. Table~\ref{tab:ctr_ablation} reports these ablation results.

\begin{table}[b]
    \centering
    \caption{ Component ablation on Electronics.}
    \label{tab:ctr_ablation}
    \resizebox{\linewidth}{!}{
    \begin{tabular}{lrrrr}
        \toprule
        \textbf{Variant}
        & \textbf{AUC}
        & \textbf{Drop}
        & \textbf{GAUC}
        & \textbf{Drop} \\
        \midrule
        \rowcolor{lightgray}\textbf{SIDRec}
        & \textbf{0.87954}
        & --
        & \textbf{0.87758}
        & -- \\

        \textit{w/o} Collaborative View
        & 0.87610
        & \textcolor{red}{0.39\%$\downarrow$}
        & 0.87372
        & \textcolor{red}{0.44\%$\downarrow$} \\

        \textit{w/o} End-to-end SID
        & 0.86223
        & \textcolor{red}{1.97\%$\downarrow$}
        & 0.86040
        & \textcolor{red}{1.96\%$\downarrow$} \\

        \textit{w/o} Both
        & 0.81627
        & \textcolor{red}{7.19\%$\downarrow$}
        & 0.81516
        & \textcolor{red}{7.11\%$\downarrow$} \\
        \bottomrule
    \end{tabular}}
\end{table}

\begin{itemize}
    \setlength{\itemsep}{2pt}
    \setlength{\parskip}{0pt}
    \item \textbf{Effectiveness of end-to-end optimization.} Removing end-to-end SID optimization produces relative AUC and GAUC drops of 1.97\% and 1.96\%, respectively. These results show the contribution of end-to-end SID optimization to ranking performance.
    \item \textbf{Contribution of collaborative fusion.} Removing collaborative information leads to relative AUC and GAUC drops of 0.39\% and 0.44\%, respectively. These results show that collaborative fusion improves the discriminative capability of the learned SID representations.
    \item \textbf{Impact of joint removal.} Removing both components produces the largest relative drops, with AUC decreasing by 7.19\% and GAUC by 7.11\%. This larger degradation supports the combined contribution of collaborative fusion and end-to-end SID optimization.
\end{itemize}
Overall, the ablation results show that both collaborative fusion and end-to-end SID optimization contribute to SIDRec's ranking performance.

\subsubsection{Collaborative Sensitivity}

\begin{figure}[h]
    \centering
    \includegraphics[width=\linewidth]{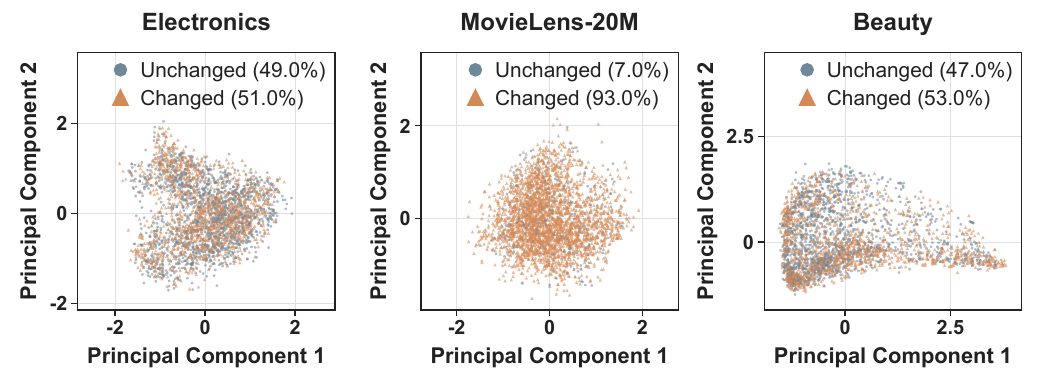}
    \caption{First-level SID assignments after shuffling collaborative embeddings across items with model parameters fixed.}
    \label{fig:collab_sid}
\end{figure}

We examine how collaborative inputs influence the discrete item identifiers. We randomly shuffle collaborative embeddings across items and recompute the first-level SID assignments while keeping model parameters fixed. As shown in Figure~\ref{fig:collab_sid}, 51.0\%, 93.0\%, and 53.0\% of items change their assignments on Electronics, MovieLens-20M, and Beauty, respectively. MovieLens-20M exhibits the highest reassignment rate. These results demonstrate that item-level collaborative information directly influences discrete SID assignments across all three datasets.

\subsubsection{SID Consistency}
Table~\ref{tab:sid_stability} examines latent representation stability and the structural consistency of first-level SID assignments.
\begin{table}[t]
\centering
\caption{
Semantic consistency of SID assignments under collaborative perturbation.}
\label{tab:sid_stability}
\setlength{\tabcolsep}{4pt}
\renewcommand{\arraystretch}{1.08}
\resizebox{\columnwidth}{!}{%
\begin{tabular}{lrrr}
\toprule
\textbf{Metric}
& \textbf{Electronics}
& \textbf{ML-20M}
& \textbf{Beauty} \\
\midrule

\rowcolor{lightgray}
\multicolumn{4}{l}{\textbf{\textit{Representation Stability}}} \\
\midrule
Latent Cosine (Changed) $\uparrow$
& 0.985 & 0.954 & 0.989 \\
Latent Cosine (Unchanged) $\uparrow$
& 0.990 & 0.979 & 0.993 \\

\midrule
\rowcolor{lightgray}
\multicolumn{4}{l}{\textbf{\textit{Structural Consistency}}} \\
\midrule
Codebook Cosine (Random) $\uparrow$
& 0.84 (0.67) & 0.87 (0.80) & 0.89 (0.69) \\
Token Cosine (Random) $\uparrow$
& 0.34 (0.00) & 0.25 (0.08) & 0.26 (0.02) \\
Partition ARI $\uparrow$
& 0.710 & 0.374 & 0.601 \\
Partition NMI $\uparrow$
& 0.852 & 0.675 & 0.835 \\
\bottomrule
\end{tabular}%
}
\end{table}
\begin{itemize}
    \setlength{\itemsep}{2pt}
    \setlength{\parskip}{0pt}
    \item \textbf{Latent representations.} Items with changed assignments retain average latent cosine similarities of 0.985, 0.954, and 0.989 across the three datasets. The corresponding values for unchanged items are 0.990, 0.979, and 0.993. These high similarities show that the latent representations remain closely aligned as discrete SID assignments change.
    \item \textbf{Code migration.} Similarities between the original and reassigned codewords exceed the random-slot baselines on all datasets. Token similarities also exceed their baselines, reaching 0.34, 0.25, and 0.26 compared with 0.00, 0.08, and 0.02. This pattern indicates that SID reassignments favor related codewords rather than arbitrary alternatives.
    \item \textbf{Partition agreement.} ARI ranges from 0.374 to 0.710, and NMI ranges from 0.675 to 0.852. MovieLens-20M has the lowest agreement under both measures, consistent with its higher reassignment rate. These results characterize how SID partitions adapt to changes in collaborative signals across datasets.
\end{itemize}
These ablation and SID analyses show that collaborative information influences discrete item representations and that end-to-end SID optimization improves ranking performance.

\subsection{SID Assignment Dynamics During Training}

\begin{figure}
    \centering
    \includegraphics[width=\linewidth]{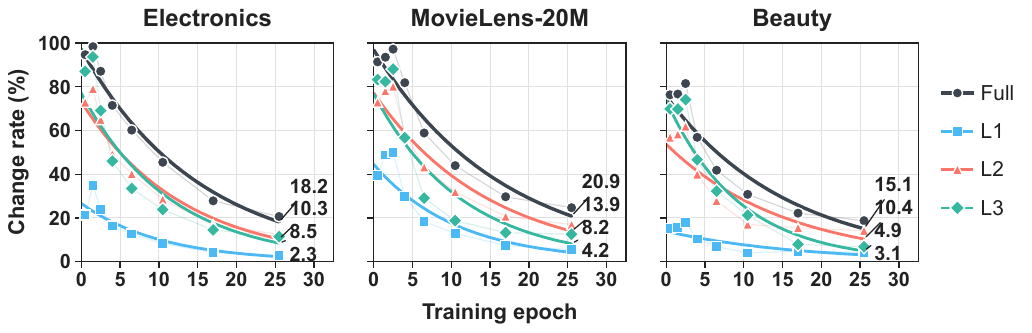}
    \caption{SID assignment dynamics during training across three datasets.}
    \label{fig:sid_dynamics}
\end{figure}

To investigate the stability of learned Semantic IDs (SIDs), we analyze their assignment dynamics throughout training across three datasets, as shown in Figure~\ref{fig:sid_dynamics}. Across all datasets, SID reassignment rates decrease as training progresses. By the final epoch, the reassignment rates of complete SIDs reach 18.2\%, 20.9\%, and 15.1\% on Electronics, MovieLens-20M, and Beauty, respectively. The corresponding first-level (L1) rates decrease to 2.3\%, 4.2\%, and 3.1\%, remaining below those at deeper levels (L2 and L3). These results show that SID assignments become progressively more stable during end-to-end training, with the first-level assignments exhibiting the lowest change rates.

\subsection{Industrial Evaluation (RQ3)}
\label{sec:exp_industry}
Finally, we validate SIDRec in real-world recommendation scenarios on our company's large-scale recommendation platform. We conduct offline evaluations across two industrial domains and three prediction tasks, including Show, CTR, and CVR. These experiments aim to verify the practical effectiveness and generalizability of SIDRec across diverse industrial ranking objectives.

\definecolor{gain0}{RGB}{248,249,248}
\definecolor{gain1}{RGB}{239,245,241}
\definecolor{gain2}{RGB}{224,238,228}
\definecolor{gain3}{RGB}{205,227,212}
\definecolor{gain4}{RGB}{181,215,191}
\definecolor{gain5}{RGB}{151,199,164}

\begin{table}[ht]
    \centering
    \caption{Offline gains over industrial baselines in percentage points. Darker green indicates larger gains within each metric. The largest gain within each domain and metric is bold.}
    \label{tab:industrial_results}
    \small
    \setlength{\tabcolsep}{3pt}
    \begin{tabular*}{\linewidth}{@{\extracolsep{\fill}}llrrr@{}}
        \toprule
        \textbf{Domain} & \textbf{Task} & $\mathbf{\Delta}$\textbf{AUC} & $\mathbf{\Delta}$\textbf{UAUC} & $\mathbf{\Delta}$\textbf{WUAUC} \\
        \midrule
        \multirow{3}{*}{\textit{Domain 1}}
        & Show & \cellcolor{gain2}+0.50 & \cellcolor{gain1}+0.62 & \cellcolor{gain2}+1.13 \\
        & CTR & \cellcolor{gain1}+0.36 & \cellcolor{gain3}+1.15 & \cellcolor{gain2}+1.05 \\
        & CVR & \cellcolor{gain5}\textbf{+1.08} & \cellcolor{gain5}\textbf{+2.57} & \cellcolor{gain5}\textbf{+2.48} \\
        \midrule
        \multirow{3}{*}{\textit{Domain 2}}
        & Show & \cellcolor{gain2}\textbf{+0.42} & \cellcolor{gain2}+0.60 & \cellcolor{gain3}+1.11 \\
        & CTR & \cellcolor{gain1}+0.08 & \cellcolor{gain0}+0.06 & \cellcolor{gain0}+0.06 \\
        & CVR & \cellcolor{gain0}+0.00 & \cellcolor{gain5}\textbf{+1.97} & \cellcolor{gain5}\textbf{+2.08} \\
        \bottomrule
    \end{tabular*}
\end{table}
Table~\ref{tab:industrial_results} reports offline gains over the corresponding baselines. UAUC and WUAUC improve in all six domain--task pairs, and AUC improves in five pairs. The three prediction objectives exhibit the following improvements.
\begin{itemize}
    \setlength{\itemsep}{2pt}
    \setlength{\parskip}{0pt}
    \item \textbf{Exposure prediction.} Show gains are consistent across the two domains. AUC improves by 0.50 and 0.42 pp, UAUC by 0.62 and 0.60 pp, and WUAUC by 1.13 and 1.11 pp. These closely matched improvements demonstrate SIDRec's effectiveness for exposure prediction in both industrial settings.
    \item \textbf{Click prediction.} Domain 1 improves by 0.36 pp in AUC, 1.15 pp in UAUC, and 1.05 pp in WUAUC. Domain 2 improves by 0.08, 0.06, and 0.06 pp, respectively. The gains across all three metrics demonstrate consistent improvements in click prediction across both domains.
    \item \textbf{Conversion prediction.} CVR achieves the largest UAUC and WUAUC gains within each domain. Domain 1 improves by 1.08 pp in AUC, 2.57 pp in UAUC, and 2.48 pp in WUAUC. Domain 2 records a change of +0.00 pp in AUC, with improvements of 1.97 pp in UAUC and 2.08 pp in WUAUC. These results show the strongest user-level improvements for conversion prediction.
\end{itemize}
In summary, SIDRec consistently improves prediction quality across both industrial domains and all three tasks, with gains in UAUC and WUAUC across all six domain--task pairs and AUC improvements in five. The particularly strong gains in conversion prediction highlight its effectiveness in capturing user-level behavioral patterns, demonstrating the practical value of integrating semantic and collaborative information into industrial recommendation models.

\section{Conclusion}

In this paper, we proposed SIDRec, a recommendation-aware semantic ID learning framework for discriminative recommendation. To address the semantic and objective mismatches in conventional SID-based methods, SIDRec integrates content and collaborative representations into SID construction and enables end-to-end optimization under downstream recommendation supervision. By aligning discrete representation learning with recommendation objectives, SIDRec captures preference-relevant information and produces more discriminative semantic identifiers. Extensive experiments on three public benchmarks demonstrate that SIDRec consistently outperforms all competing methods across both AUC and GAUC, achieving up to 5.15\% relative improvements over the strongest baselines. Further ablation studies confirm the effectiveness of incorporating collaborative semantics and jointly optimizing SID assignments with recommendation objectives. Overall, these findings highlight the importance of recommendation-aware semantic ID learning and establish SIDRec as an effective framework for discriminative recommendation.

\bibliographystyle{ACM-Reference-Format}
\balance
\bibliography{citation}

\end{document}